# VISUALIZATION AND ANALYSIS OF GEOGRAPHICAL CRIME PATTERNS USING FORMAL CONCEPT ANALYSIS

Kester, Quist-Aphetsi, MIEEE, Lecturer Faculty of Informatics, Ghana Technology University College, PMB 100 Accra North, Ghana
Phone Contact +233 209822141 Email: kquist-aphetsi@gtuc.edu.gh / kquist@ieee.org

## Abstract

There are challenges faced in today's world in terms of crime analysis when it comes to graphical visualization of crime patterns. Geographical representation of crime scenes and crime types become very important in gathering intelligence about crimes. This provides a very dynamic and easy way of monitoring criminal activities and analyzing them as well as producing effective countermeasures and preventive measures in solving them. But we need effective computer tools and intelligent systems that are automated to analyze and interpret criminal data in real time effectively and efficiently. These current computer systems should have the capability of providing intelligence from raw data and creating a visual graph which will make it easy for new concepts to be built and generated from crime data in order to solve, understand and analyze crime patterns easily.

This paper proposes a new method of visualizing and analyzing crime patterns based on geographical crime data by using Formal Concept Analysis, or Galois Lattices, a data analysis technique grounded on Lattice Theory and Propositional Calculus. This method considered the set of common and distinct attributes of crimes in such a way that categorization are done based on related crime types.

This will help in building a more defined and conceptual systems for analysis of geographical crime data that can easily be visualized and intelligently analyzed by computer systems.

**Keywords: FCA, crime patters, visualization, analyses**

## Introduction

Virtually all societies in the modern world are troubled by crime. While crime rates vary enormously from one country to another and from one region to another, criminal behavior remains a cause for concern amongst most members of the public. Solving crimes has been the prerogative of the criminal justice and law enforcement specialists. With the increasing use of the computerized systems to track crimes, computer data analysts have started helping the law enforcement officers and detectives to speed up the process of solving crimes. The most efficient and effective way of fighting crime today cannot be resourceful without geographical profiling. Criminal activities have become very complex in such a way that rapid monitory can only be achieved by using intelligent systems with geographical components.

Geographic profiling is growing in popularity and, combined with offender profiling, can be a helpful tool in the investigation of serial crime. It is a criminal investigative methodology that analyzes the locations of a connected series of crimes to determine the most probable area of offender residence. By incorporating both qualitative and quantitative methods, it assists in understanding spatial behavior of an offender and focusing the investigation to a smaller area of the community. Typically used in cases of serial murder or rape (but also arson, bombing, robbery, and other crimes), the technique helps police detectives prioritize information in large-scale major crime investigations that often involve hundreds or thousands of suspects and tips.[1]

In addition to determining the offender's most likely area of residence, an understanding of the spatial pattern of a crime series and the characteristics of the crime sites can tell investigators other useful information, such as whether the crime was opportunistic and the degree of offender familiarity with the crime location. This is based on the connection between an offender's hunting behavior and his or her non-criminal life. [1] [2]

A criminal pattern analysis is very crucial in combating crime. Computer systems have to be engaged in order to gather and interpret intelligence so as to control the criminal environment as well as influence effective decision making as in figure 1 below.

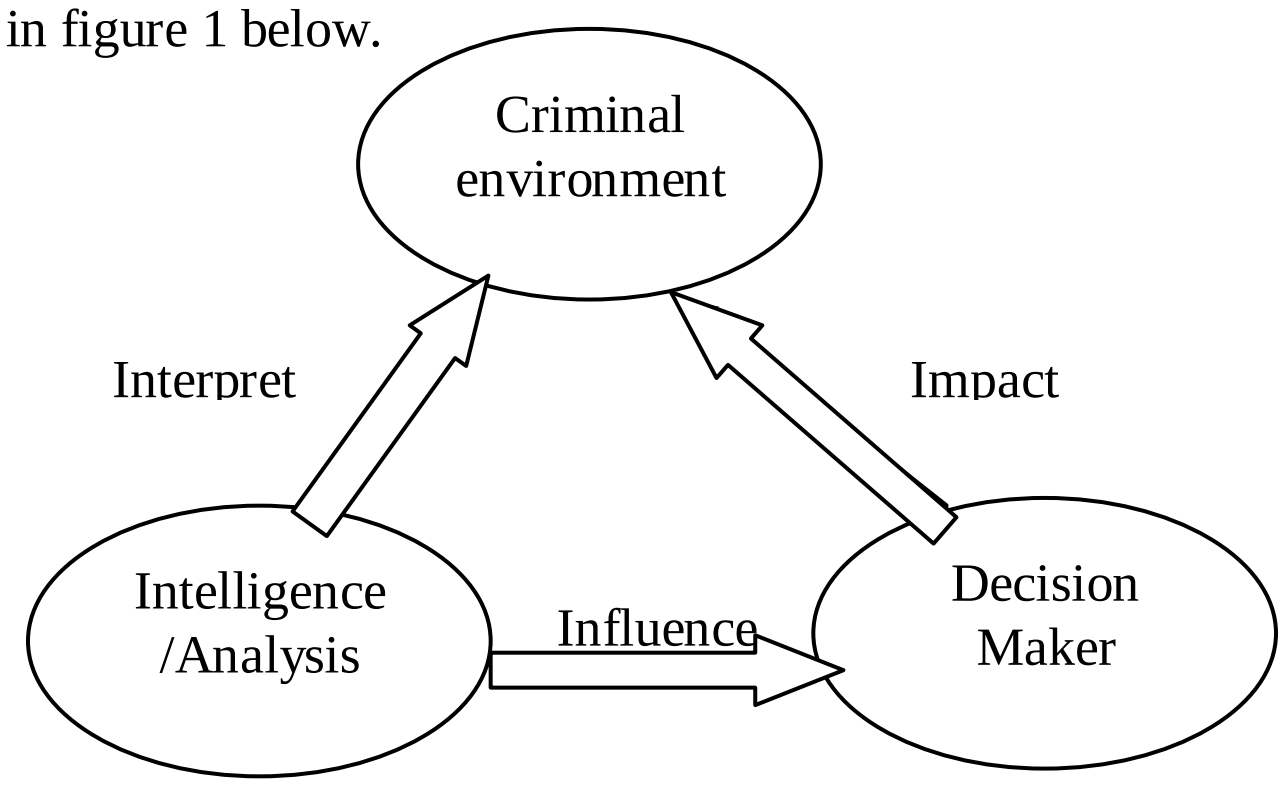

**Figure 1. Pattern analysis theory**

Formal Concept Analysis (FCA) is a mathematical theory of data analysis using formal contexts and concept lattices. [3]. It is a principled way of deriving a concept hierarchy or for-mal ontology from a collection of objects and their properties. Each concept in the hierarchy represents the set of objects sharing the same values for a certain set of properties; and each sub-concept in the hierarchy contains a subset of the objects in the concepts above it [4]. Its range of applications can be found in information and knowledge processing including visualization, data analysis and knowledge management

The aim of this paper is to propose a new way of analyzing crime patterns by combining Formal concept analysis, or Galois Lattices, a data analysis technique grounded on Lattice theory and propositional Calculus, and geographic information science to discover the patterns within criminal data. This method considered the set of common and distinct attributes of crimes in such a way that categorization are done based on related crime types. At the end this will help in building a more defined and conceptual systems for analysis of geographical crime data that can easily be visualized and intelligently analyzed by computer systems.

The organization of this paper is as follows, section II of this paper provides details of the related work in the domain of crime pattern analysis and application of FCA in the field of computer science. Section III proposes how the FCA will be used to classify and analyze crime. Section IV provides application and results: provides the use of FCA crime analysis. The last section of this paper, section V, Concludes the paper.

# Related Works

Crime activities are geospatial phenomena and as such are geospatially, thematically and temporally correlated. Thus, crime datasets must be interpreted and analyzed in conjunction with various factors that can contribute to the formulation of crime. Discovering these correlations allows a deeper insight into the complex nature of criminal behavior.

Phillips, P and Lee, I. introduced a graph based dataset representation that allows them to mine a set of datasets for correlation. They demonstrated their approach with real crime datasets and provided a comparison with other techniques. [5]

Data mining can be used to model crime detection problems. Here we look at use of clustering algorithm for a data mining approach to help detect the crimes patterns and speed up the process of solving crime. Shyam Varan Nath presented a paper that looked at k-means clustering with some enhancements to aid in the process of identification of crime patterns. They applied these techniques to real crime data from a sheriff's office and validated their results and further they used semi-supervised learning technique for knowledge discovery from the crime records to help increase the predictive accuracy. They developed a weighting scheme for attributes to deal with limitations of various out of the box clustering tools and techniques. Their framework worked with geospatial plot of crime data. [6]

P. Rogerson and Y. Sun described a new procedure for detecting changes over time in the spatial pattern of point events, combining the nearest neighbor statistic and cumulative sum methods. The method results in the rapid detection of deviations from expected geographic patterns. The method was illustrated using 1996 arson data from the Buffalo, NY, Police Department. [7]

The appearance of patterns could be found in different modalities of a domain, where the different modalities refer to the data sources that constitute different aspects of a domain. Particularly, the domain that refers to crime and the different modalities refer to the different data sources within the crime domain such as offender data, weapon data, etc. In addition, patterns also exist in different levels of granularity for each modality. In order to have a thorough understanding a domain, it is important to reveal hidden patterns through the data explorations at different levels of granularity and for each modality. Therefore, Yee Ling Boo and Alahakoon, D presented a new model for identifying patterns that exist in different levels of granularity for different modes of crime data. A hierarchical clustering approach - growing self organizing maps (GSOM) has been deployed. The model was further enhanced with experiments that exhibit the significance of exploring data at different granularities. [8]

Formal concept analysis (FCA) is a method of data analysis with growing popularity across various domains .FCA analyzes data which describe relationship between a particular set of objects and a particular set of attributes. Such data commonly appear in many areas of human activities. FCA produces two kinds of output from the input data. The first is a concept lattice. A concept lattice is a collection of formal concepts in the data which are hierarchically ordered by a subconcept-super concept relation. Formal concepts are particular clusters which represent natural human-like concepts such as “organism living in water”, “car with all wheel drive system”, “number divisible by 3 and 4”,etc. The second output of FCA is a collection of so-called attribute implications. An attribute implication describes a particular dependency which is valid in the data such as “every number divisible by 3 and 4 is divisible by 6”,“every respondent with age over 60 is retired”, etc.[9]

In the field of software engineering and computer science, Formal Concept Analysis (FCA) has generally been applied to support software maintenance activities — the refactoring or medication of existing code — and to the identification of object-oriented (OO) structures.

Risk Matrix is a matrix that is used during Risk Assessment to define the various levels of risk as the product of the harm probability categories and harm severity categories. This is a simple mechanism to increase visibility of risks and assist management decision making. Risk matrix is presented for use in identifying and assessing project risks quickly and cost effectively. But how risks are related and visualized at various phases becomes very difficult to be seen as the activity becomes many and complex with software projects. QA kester proposed a new method of evaluation, analysis and visualization of the Assessment Evaluation of Risks Matrix in software engineering based on Formal Concept Analysis, or Galois Lattices, a data analysis technique grounded on Lattice Theory and Propositional Calculus. The method has helped in building a more defined and conceptual systems for Evaluation of risk Levels that can easily be visualized in software engineering projects. [10]

There is also a body of literature reporting the application of FCA to the identification and maintenance of class hierarchies in database schemata [11, 12, 13]. And also Simon Andrew sand Simon Polovina presented an outline of a process by which operational Software requirements specification can be written for Formal Concept Analysis (FCA). The Z notation was issued to specify the FCA model and the formal operations on it. They conceive a novel approach where by key features of Z and FCA can be integrated and put to work in contemporary software development, thus promoting operational specification as a useful application of conceptual structures. [14]

Module and Object Identification with FCA: Sahraoui et al. [15] present a method that extracts objects from procedural code using FCA. Important concepts are looked for in the resultant lattices using heuristics. Another approach compares the object identification capability of clustering and FCA techniques [16]. A few heuristics are described to interpret the concepts of the lattice. An approach to transform a COBOL legacy system to a distributed component system is proposed by Canfora et al. [17]. Siff and Reps explore the relationship between program variables and procedures through FCA for restructuring of C programs [18].

Modern police organizations and intelligence services are adopting the use of FCA in crime pattern analysis for tracking down criminal suspects through the integration of heterogeneous data sources and visualizing this information so that a human expert can gain insight in the data [19].

# Methodology

Formal Concept Analysis, or Galois Lattices, is a data analysis technique grounded on Lattice Theory and Propositional Calculus. This paper proposes a new method based on FCA method and considered the set of objects and attributes of crime patterns in such a way that categorization are done based on crime types. The crimes are first grouped into various types with their places of occurrence.

In FCA a formal context consists of a set of objects, G, a set of attributes, M, and a relation between G and M, $I \subseteq G \times M$. A formal concept is a pair (A,B) where $A \subseteq G$ and $B \subseteq M$. Every object in A has every attribute in B. For every object in G that is not in A, there is an attribute in B that that object does not have. For every attribute in M that is not in B there is an object in A that does not have that attribute. A is called the extent of the concept and B is called the intent of the concept.

If $g \in A$ and $m \in B$ then $(g,m) \in I$ ,or gIm.
A formal context is a tripel (G,M,I), where
• G is a set of objects,
• M is a set of attributes
•and I is a relation between G and M.
• $(g,m) \in I$ is read as „object g has attribute m".

For $A \subseteq G$, we define
$A' := \{m \in M \mid \forall g \in A : (g,m) \in I\}$.
For $B \subseteq M$, we define dually
$B' := \{g \in G \mid \forall m \in B : (g,m) \in I\}$.

For A, $A_1$, $A_2 \subseteq G$ holds:
• $A_1 \subseteq A_2 \Rightarrow A'_2 \subseteq A'_1$
• $A_1 \subseteq A''$
• $A' = A'''$

For B, B1, B2 $\subseteq$ M holds:
• $B_1 \subseteq B_2 \Rightarrow B'_2 \subseteq B'_1$
• $B \subseteq B''$
• $B' = B'''$

A formal concept is a pair (A, B) where
• A is a set of objects (the extent of the concept),
• B is a set of attributes (the intent of the concept),
•$A' = B$ and $B' = A$.

The concept lattice of a formal context (G, M, I) is the set of all formal concepts of (G, M, I), together with the partial order
$(A_1, B_1) \le (A_2, B_2) :\Leftrightarrow A_1 \subseteq A_2$ $(\Leftrightarrow B_1 \supseteq B_2)$.

The concept lattice is denoted by $\mathfrak{B}(G,M,I)$ .
• Theorem: The concept lattice is a lattice, i.e. for two concepts $(A_1, B_1)$ and $(A_2, B_2)$, there is always
•a greatest common subconcept: $(A_1 \cap A_2, (B_1 \cup B_2)'')$
•and a least common superconcept: $((A_1 \cup A_2)'', B_1 \cap B_2)$

More general, it is even a complete lattice, i.e. the greatest common subconcept and the least common superconcept exist for all (finite and infinite) sets of concepts.
Corollary: The set of all concept intents of a formal context is a closure system. The corresponding closure operator is h(X):= X``.
An implication X→Y holds in a context, if every object having all attributes in X also has all attributes in Y.
Def.: Let X ⊆M. The attributes in X are independent, if there are no trivial dependencies between them.

## Applications and Results

We consider crime types such that P belong to the set of crimes and P1, P2…Pn are subsets of P. Let P1= murder P2= robbery P3= theft P4= assault P5= burglary P6= Money Laundering P7= Kidnapping and finally P8= Theft. Where the objects are the geographical locations (A, B, C, D, E and F) indicated on the map below in figure 2.

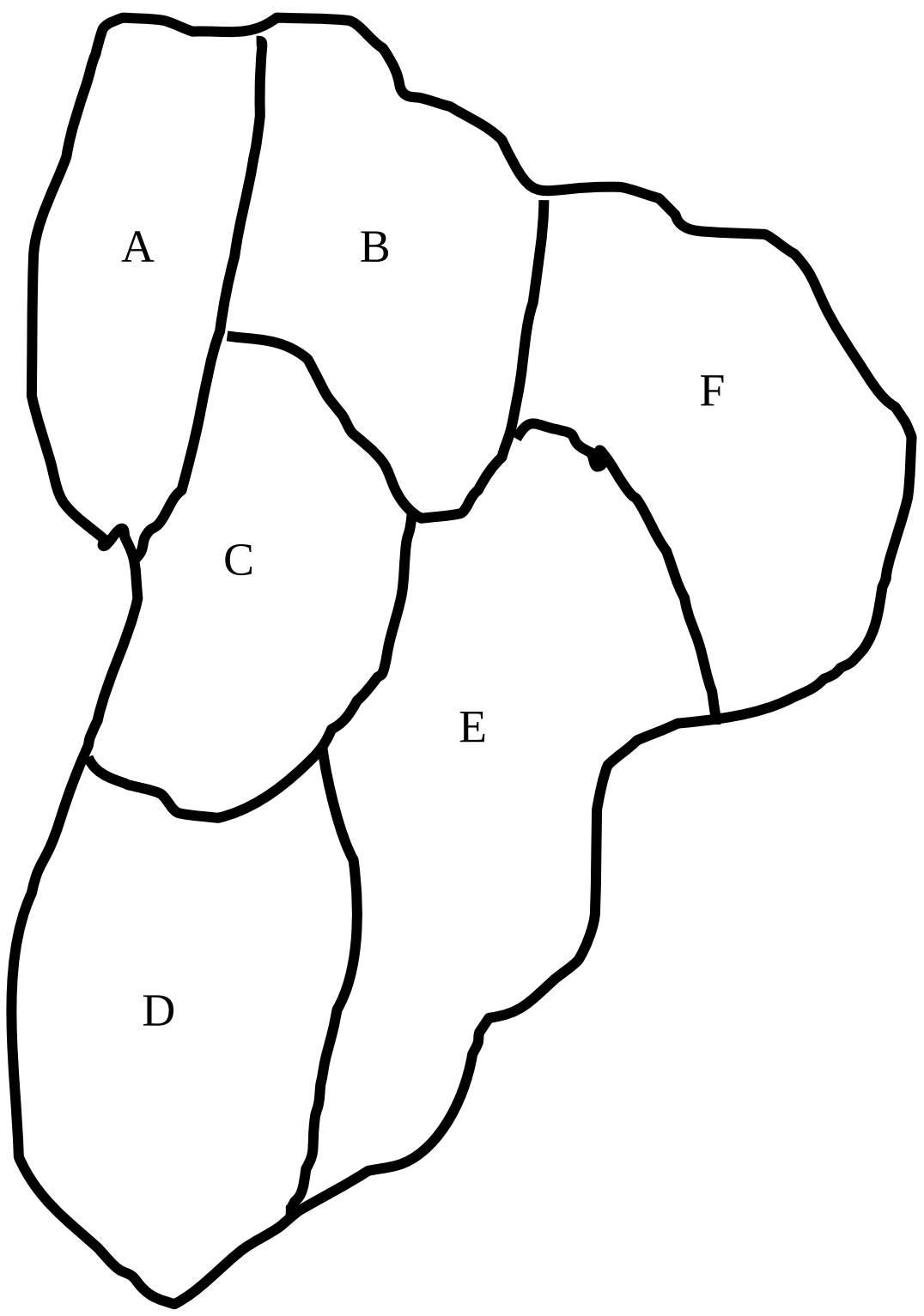

**Figure 2. Geographical locations of a map**

**Table 1. Geographical locations cross crime types**

| | P1 | P2 | P3 | P4 | P5 | P6 | P7 | P8 |
|---|---|---|---|---|---|---|---|---|
| A | x | | x | | | | x | |
| B | | x | x | | x | | | x |
| C | | | | x | x | x | | |
| D | x | | x | | x | | x | |
| E | x | x | | | | x | | |
| F | | x | | x | x | | x | x |

The table one was transformed into xml code below and run in Galicia, Galois Lattice builder Software to produce figure 3 and figure 4.

```
<?xml version="1.0" encoding="UTF-8"?>
<BIN name="Crime patternContext.slf" nbObj="6" nbAtt="8" type="BinaryRelation">
  <OBJS>
    <OBJ id="0">A</OBJ>
    <OBJ id="1">B</OBJ>
    <OBJ id="2">C</OBJ>
    <OBJ id="3">D</OBJ>
    <OBJ id="4">E</OBJ>
    <OBJ id="5">F</OBJ>
  </OBJS>
  <ATTS>
    <ATT id="0">P1</ATT>
    <ATT id="1">P2</ATT>
    <ATT id="2">P3</ATT>
    <ATT id="3">P4</ATT>
    <ATT id="4">P5</ATT>
    <ATT id="5">P6</ATT>
    <ATT id="6">P7</ATT>
    <ATT id="7">P8</ATT>
  </ATTS>
  <RELS>
    <REL idObj="0" idAtt="0" />
    <REL idObj="0" idAtt="2" />
    <REL idObj="0" idAtt="6" />
    <REL idObj="1" idAtt="1" />
    <REL idObj="1" idAtt="2" />
    <REL idObj="1" idAtt="4" />
    <REL idObj="1" idAtt="7" />
    <REL idObj="2" idAtt="3" />
    <REL idObj="2" idAtt="4" />
    <REL idObj="2" idAtt="5" />
    <REL idObj="3" idAtt="0" />
    <REL idObj="3" idAtt="2" />
    <REL idObj="3" idAtt="4" />
    <REL idObj="3" idAtt="6" />
    <REL idObj="4" idAtt="0" />
    <REL idObj="4" idAtt="1" />
```

```
    <REL idObj="4" idAtt="5" />
    <REL idObj="5" idAtt="1" />
    <REL idObj="5" idAtt="3" />
    <REL idObj="5" idAtt="4" />
    <REL idObj="5" idAtt="6" />
    <REL idObj="5" idAtt="7" />
  </RELS>
</BIN>
```

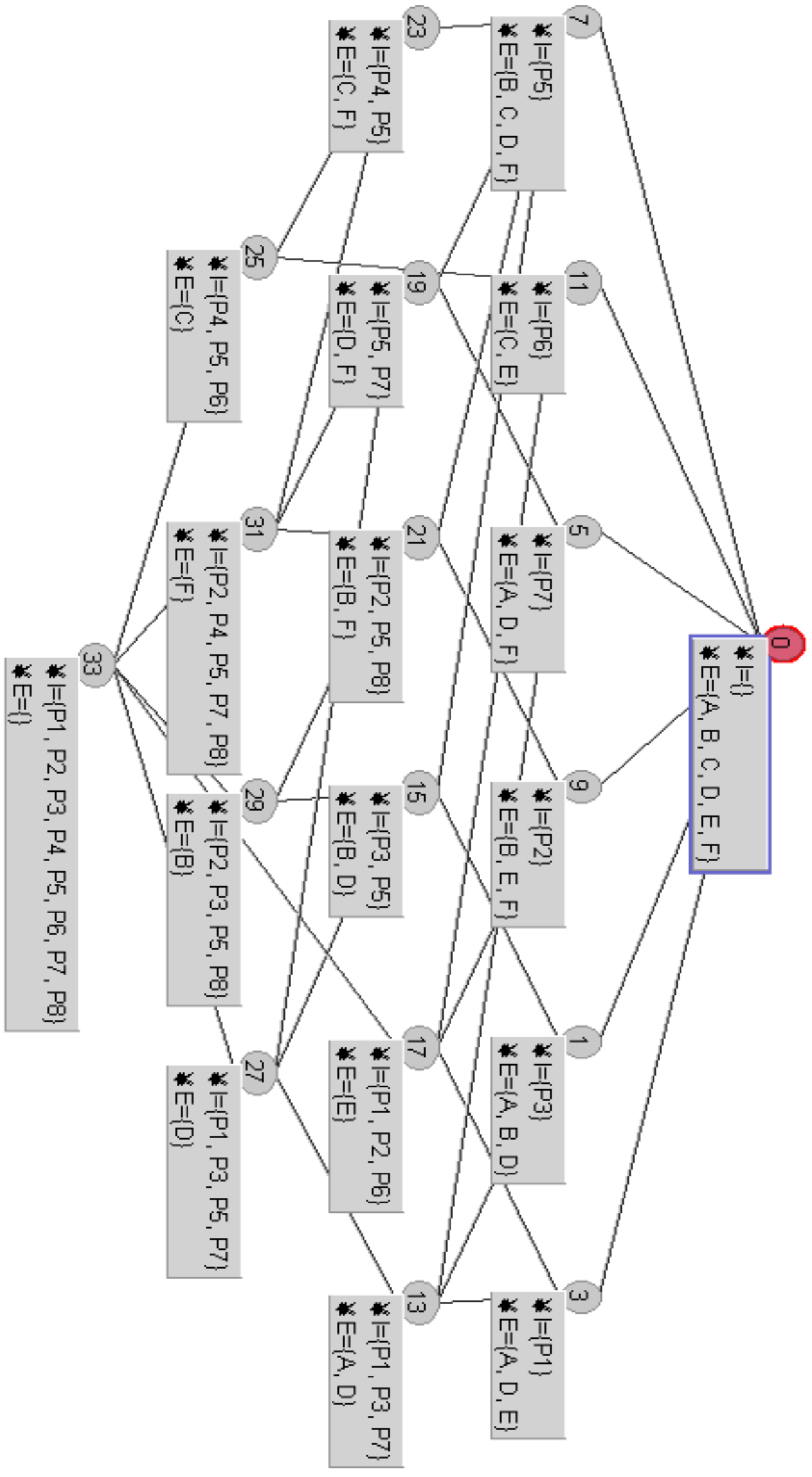

**Figure 3. Galois lattices of intents and extents**

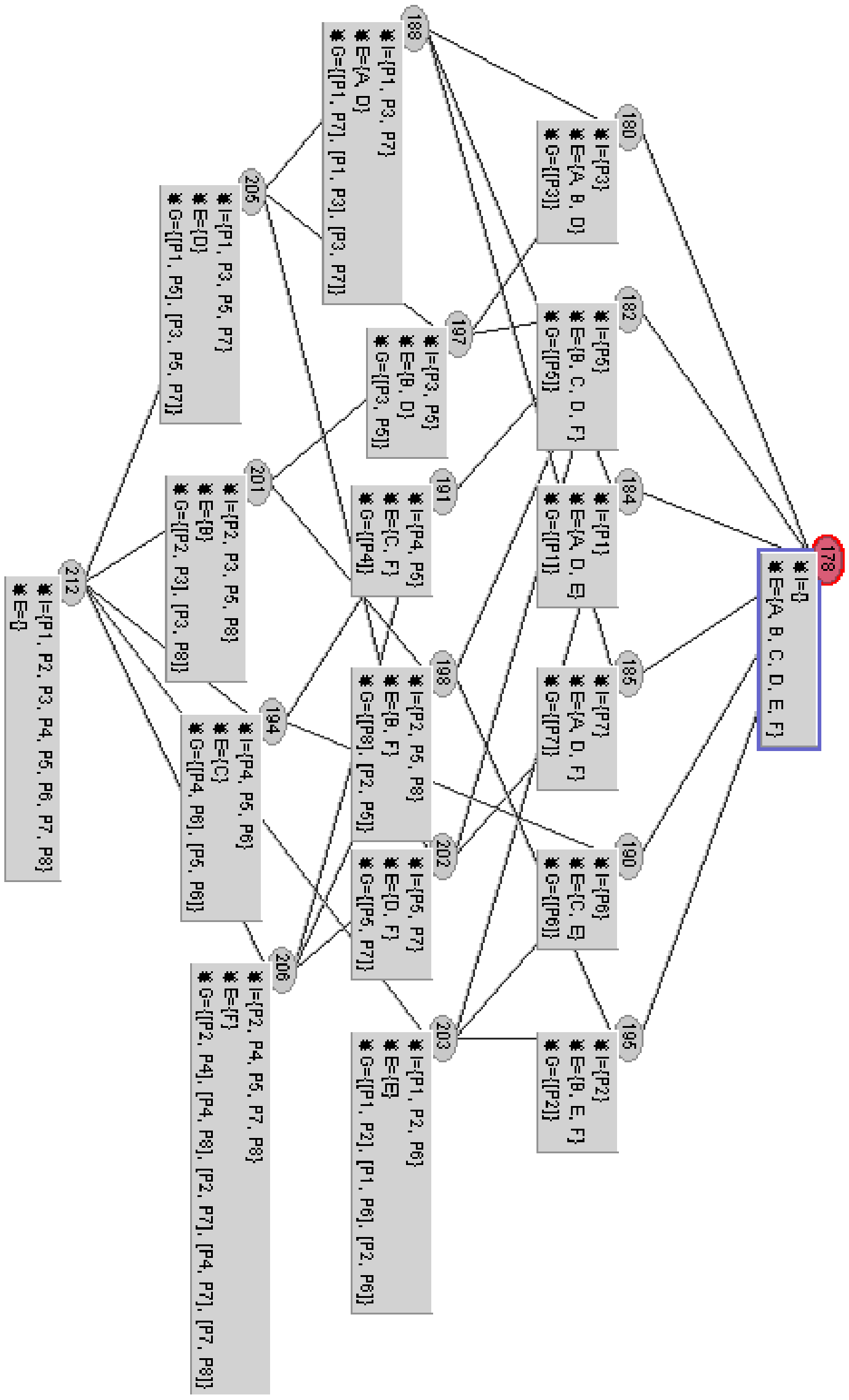

**Figure 4. Figure 4 Galois lattice of (G, M, I)**

The concept lattice of a formal context (G, M, I) as shown in figure 4, is the set of all formal concepts of (G, M, I), together with the partial order (A1, B1) ≤ (A2, B2): ⇔ A1 ⊆ A2 (⇔ B1 ⊇ B2). It is clearly seen visually from figure 3 as well as figure 4 that the attributes belonging to concept 11and 23 in figure 3, and concept 190 and 191 in figure 4 are subsumed by concept 25 and 194 respectively. This means that location C has a related crime type with location E which is P6 and location C also has a relation with location F in terms of crime type P4. Hence context 194 has intent I= {P4, P5, P6} and Extent of E= {c} and of G= {[P4, P6] [P5, P6]}

# Conclusion

A formal concept analysis was used to analyze crime patterns and visualized relationships between the occurrences of various crimes within different geographical areas. This method considered the set of common and distinct attributes of crimes in such a way that categorization was done based on related crime types. This will help in building a more defined and conceptual systems for analysis of geographical crime data that can easily be visualized and intelligently analyzed by computer systems.

# Biographies

**QUIST-APHETSI KESTER, MIEEE:** is a global award winner 2010 (First place Winner with Gold), in Canada Toronto, of the NSBE's Consulting Design Olympiad Awards and has been recognized as a Global Consulting Design Engineer. He is a law student at the University of London, UK. He is a PhD student in Computer Science. The PhD program is in collaboration between the AWBC/ Canada and the Department of Computer Science and Information Technology (DCSIT), University of Cape Coast. He had a Master of Software Engineering degree from the OUM, Malaysia and BSC in Physics from the University of Cape Coast-UCC Ghana.

He has worked in various capacities as a peer reviewer for IEEE ICAST Conference, IET-Software Journal, lecturer, Head of Digital Forensic Laboratory Department at the Ghana Technology University and Head of Computer science department. He is currently a lecturer at the Ghana Technology University College and He may be reached at kquist-aphetsi@gtuc.edu.gh or kquist@ieee.org.